# Spiraling Beam Illumination Uniformity on Heavy Ion Fusion Target

T. Kurosaki[a], S. Kawata[a], K. Noguchi[a], S. Koseki[a], D. Barada[a], Y. Y. Ma[a], A. I. Ogoyski[b], J. J. Barnard[c] and B. G. Logan[c]

[a]Graduate School of Eng., Utsunomiya University, 7-1-2 Yohtoh, Utsunomiya 321-8585, Japan
[b]Department of Physics, Varna Technical University, Varna 9010, Bulgaria
[c]Lawrence Berkeley National Lab. and Virtual National Lab. for Heavy Ion Fusion, Berkeley, California 94720, USA

A few % wobbling-beam illumination nonuniformity is realized in heavy ion inertial confinement fusion (HIF) by a spiraling beam axis motion in the paper. So far the wobbling heavy ion beam (HIB) illumination was proposed to realize a uniform implosion in HIF. However, the initial imprint of the wobbling HIBs was a serious problem and introduces a large unacceptable energy deposition nonuniformity. In the wobbling HIBs illumination, the illumination nonuniformity oscillates in time and space. The oscillating-HIB energy deposition may contribute to the reduction of the HIBs' illumination nonuniformity. The wobbling HIBs can be generated in HIB accelerators and the oscillating frequency may be several 100MHz~1GHz. Three-dimensional HIBs illumination computations presented here show that the few % wobbling HIBs illumination nonuniformity oscillates successfully with the same wobbling HIBs frequency.

PACS numbers: 52.58.Hm, 41.85.-p

# I. INTRODUCTION

Heavy ion beams (HIB) have preferable features in inertial confinement fusion (ICF),[1] high energy density physics,[2-4] also ion canter therapy, etc: a HIB pulse shape is controlled precisely to fit various requirements, a HIB axis is also precisely controllable, a HIB generation energy efficiency is 30~40%, a HIB particle energy deposition is almost classical, and the deposition profile is well defined. Especially HIBs deposit their main energy in a deep area of the target material.

One of important issues in ICF is the fuel target implosion uniformity; a sufficiently uniform driver energy deposition is required[1, 5, 6]. Therefore, so far a dynamic stabilization method for the Rayleigh-Taylor (R-T) instability in the target implosion has been proposed and studied.[7-12] On the other hand, the HIB axis controllability provides a unique tool to smooth the HIB energy deposition nonuniformity, and can introduce wobbling or axis-oscillating HIBs.[13, 14] The wobbling HIBs may contribute to reduce the HIBs' illumination nonuniformity. In addition, our previous work showed that the R-T instability growth can be reduced by the sinusoidally oscillating acceleration in time and space[15, 16], if the wobbling HIBs provides the oscillating acceleration required and at the same time the wobblers' illumination is sufficiently uniform. In this sense, the wobbling HIBs may have a potential to realize a uniform energy deposition. So far the wobbling HIB has been studied to smooth the HIB illumination for a cylindrical target.[3, 17-19]

Detailed studies of three-dimensional HIBs illumination show that HIBs illumination uniformity depends strongly on the HIBs illumination scheme.[20, 21] For the wobbling HIBs illumination (see Figs. 1 and 2), Ref. 21 showed a sufficiently small nonuniformity for circularly wobbling HIBs in a steady state, and the nonuniformity was evaluated after the wobbling HIBs energy deposition becomes steady.[21] However, the wobbling HIBs illumination is time-dependent. We found that in particular the initial HIBs illumination nonuniformity becomes large and is not acceptable in ICF. In this paper we present another HIBs illumination scheme for a direct drive spherical target. In the HIBs illumination scheme proposed here, each HIB axis has a spiraling trajectory (see Fig. 3), so that the time-dependent HIBs illumination realizes a sufficiently low nonuniformity (<3.57%) from the initial time to the HIBs pulse termination.

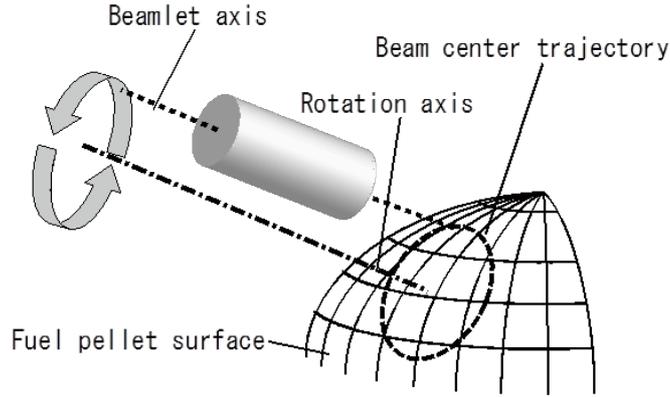

Fig. 1　The schematic diagram of wobbling HIB illumination.

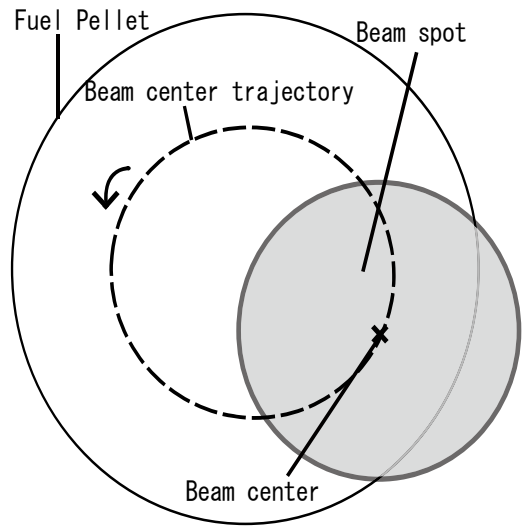

Fig. 2　A wobbling beam illumination on a direct-driven spherical target.

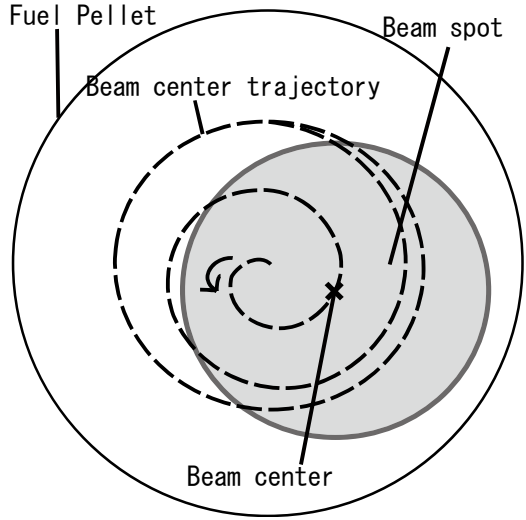

Fig. 3 Spiraling HIB. The Spiraling beam axis motion provides a sufficiently low HIBs illumination nonuniformity during the HIBs pulse duration.

## II. UNIFORMITY EVALUATION METHOD

In our studies we employ $Pb^+$ ion HIBs with the mean particle energy of 8 GeV. The beam radius at the entrance of a reactor chamber wall $R_{en}$ is 35 mm (see Fig. 4), the reactor chamber radius $R_{ch}$ is 3 m. The beam particle density distribution is the Gaussian one. The longitudinal temperature of HIB ions is 100 MeV with the Maxwell distribution. The beam transverse emittance is 3.2mm mrad, from which the focal spot radius $R_f$ and $f$ (see Fig. 4) are obtained. The target temperature increases linearly during the time of a HIB pulse deposition from 0.025 eV to 300 eV in our study. We employ an Al monolayer pellet target structure with a 4.0 mm external radius. In our study of the HIBs illumination uniformity, we use the OK code,[22] in which the detailed ion energy stopping power is computed including the beam temperature. The 32-HIBs rotation axes positions are given as presented in Ref. 23. The HIBs illumination nonuniformity is evaluated by the global *rms*, including also the Bragg peak effect in the energy deposition profile in the target radial direction.[20] The mode analyses are also performed to find the dominant mode of the illumination nonuniformity.[20] The stopping power of a target is the sum of the energy deposited in target nuclei, target bound and free electrons, and target ions:[24-26] $E_{Stop} = E_{Nucle} + E_{Bound-e} + E_{Free-e} + E_{Free-i}$, where $E_{Stop}$ is the deposition energy in the target, $E_{Nucle}$ is the deposition energy by the nucleus scattering, $E_{Bound-e}$ is by the bound electron, $E_{Free-e}$ is by free electron and $E_{Free-i}$ is by the target ion.

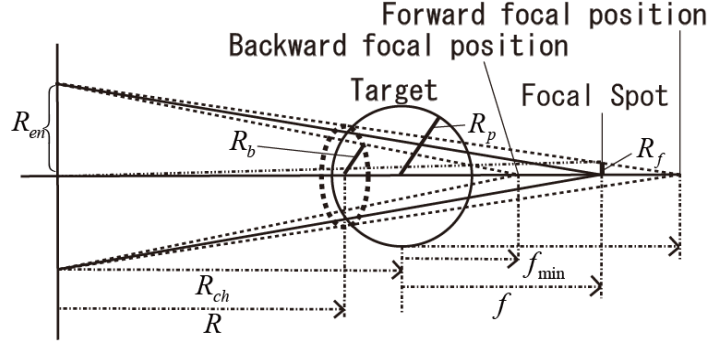

Fig. 4  A HIB illuminating on a fuel target. $Pb^+$ ion HIBs are employed with the mean particle energy of 8 GeV. The beam radius at the entrance of a reactor chamber wall $R_{en}$ is 35 mm, and the reactor chamber radius $R_{ch}$ is 5 m; $R_b$ is the beam radius on a target.

In HIB ICF the Bragg peak area of the HIB energy deposition is most important for target implosion. We employ the relative *rms* nonuniformity including the Bragg peak effect, that is, the energy deposition profile: $\sigma_i^{rms} = \frac{1}{\langle E \rangle_i} \sqrt{\sum_j^{n_\theta} \sum_k^{n_\phi} (\langle E \rangle_i - E_{i,j,k})^2 / n_\theta n_\phi}$, $\langle \sigma_{rms} \rangle = \sum_i^{n_r} w_i \sigma_i^{rms}$, $w_i = E_i/E$. In this paper each HIB is divided into many beamlets and the spherical target Al shell is divided by fine meshes in the radial *i*-direction, the polar angle *j*-direction and the azimuthal angle *k*-direction.[20, 22] Here $\langle \sigma_{rms} \rangle$ is the global *rms* nonuniformity, $\sigma_i^{rms}$ is the *rms* nonuniformity on the *i*-th surface of deposition, $w_i$ is the weight function in order to include the Bragg peak effect of the deposition profile, $n_r$, $n_\theta$ and $n_\phi$ are the total mesh numbers in each direction of the spherical coordinate, $\langle E \rangle_i$ is the mean deposition energy on the $i$-th surface, $E_i$ is the total deposition energy in the $i$-th shell, and $E$ is the total deposition energy.

We also performed mode analyses by the spherical harmonic function $Y_n^m(\theta,\phi)$: $S_n^m = \frac{1}{4\pi} \int_0^\pi d\theta \int_0^{2\pi} d\phi \sin\theta E(\theta,\phi) Y_n^m(\theta,\phi)$. Here $S_n^m$ is an amplitude of energy spectrum, $n$ and $m$ are the mode numbers, and $\theta$ and $\phi$ are the polar and azimuthal angles, respectively. $E(\theta,\phi)$ is the HIB deposition energy summarized over the radial direction at each $(\theta,\phi)$ mesh point of a target. The summation of the energy spectrum amplitude is normalized to be 1.0 in our study.

## III. SPIRALING HEAVY ION BEAM ILLUMINATION UNIFORMITY

So far the wobbling HIBs illumination on a spherical fuel target has been considered for the time averaged illumination.[21] However, it was found in this paper that the initial imprint of the rotating HIBs is serious and introduces a large illumination nonuniformity. In order to reduce the HIBs illumination nonuniformity from the initial time to the beam pulse end, in this paper we propose that each HIB axis has a spiral trajectory as shown in Fig. 3 especially in the initial few rotations. After the initial few spiral rotations, the HIB axes trajectories approach to circles.

When we do not employ the spiraling trajectory as shown in Fig. 2, the illumination nonuniformity history shows an unacceptable large nonuniformity during the first few rotations as presented in Fig. 5 (the solid line). In this case, the direct drive target radius is 4mm, the energy deposition layer consists of Al, and we employ Pb$^+$ ion HIBs with the mean energy 8GeV, as shown above. The HIB axis rotation radius is 2mm and the HIB radius is 3mm. When the spiraling HIBs are employed for the first two rotations, the HIBs illumination uniformity is drastically improved as shown in Fig. 5 (the dotted curve). In Fig. 5 the time is normalized by the wobbling beam axis rotation time $\tau$.

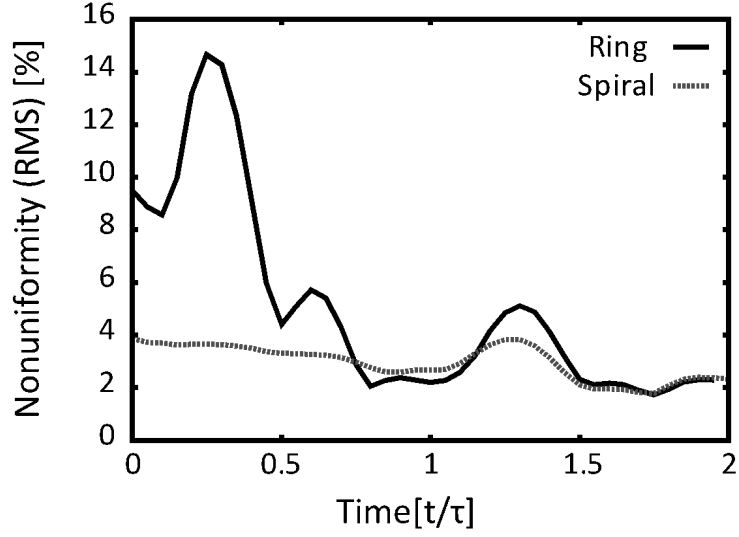

Fig. 5 Histories of energy deposition nonuniformities for the circular moving axes (solid line) and for the spiraling HIBs. The spiraling HIB-axis motion realizes a rather low HIBs illumination nonuniformity.

Figure 6 shows the history of the spiraling HIBs illumination nonuniformity as well as the HIB particle illumination loss history. Here, we define the loss as the percentage of particles that do not hit the target. By the spiraling trajectory of the HIB axis, a sufficiently small nonuniformity is successfully realized, and a small part of the HIBs particles do not hit the target in order to obtain the illumination uniformity (see Fig. 6 (the dotted line)). Figure 7 shows an energy spectrum at $t = 1.3\tau$, at which time the HIBs illumination nonuniformity has a local-peak value as shown in Fig. 5 (the black line). In Fig. 7, ($n, m$) are the polar and azimuthal mode numbers, and $S_n^m$ is the amplitude of the spectrum, respectively. If the deposition energy distributed is prefectly spherically symmetric, the amplitude of the spectrum is 1.0 in the mode

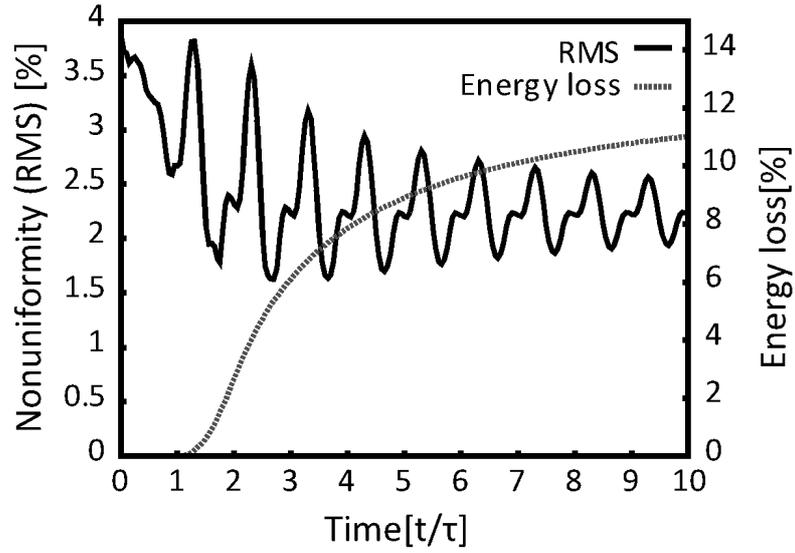

Fig. 6  Histories of the spiraling HIBs illumination nonuniformity (the solid line) and the illumination loss (the dotted line). The spiraling HIBs provide a sufficiently uniform deposition uniformity in HIF.

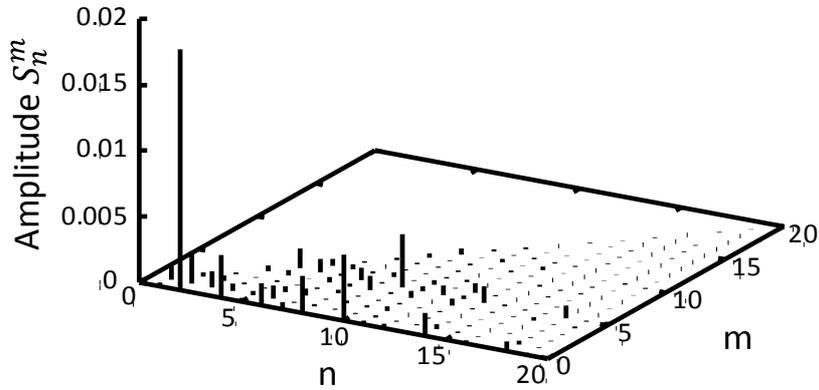

Fig. 7  The energy spectrum at $t = 1.3\tau$ of the spiraling HIBs. The mode (n, m)=(2, 0) is dominant throughout the spiral HIBs illumination.

($n, m$) = (0, 0) in our study. For this reason, the amplitude of the mode ($n, m$) = (0, 0) becomes large, nearly 1.0. In this paper the amplitude of the spectrum the mode ($n, m$) = (0, 0) is not displayed. As a result, the amplitude of spectrum mode ($n, m$) = (2, 0) is the largest mode in Fig. 7, and the mode ($n, m$)=(2, 0) is dominant throughout the HIBs illumination. Figure 8 shows the amplitude of the mode (2, 0) versus time, and Fig. 9 presents the spectrum of the mode (2, 0) amplitude in its frequency space. In Fig. 9 $f_{wb}$ shows the wobbling HIBs rotation frequency.

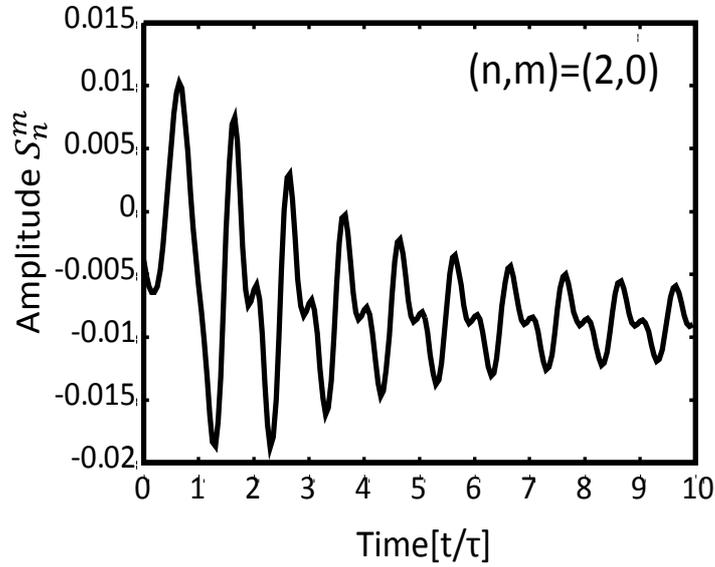

Fig. 8　The mode (2, 0) amplitude of HIBs energy deposition nonuniformity versus time.

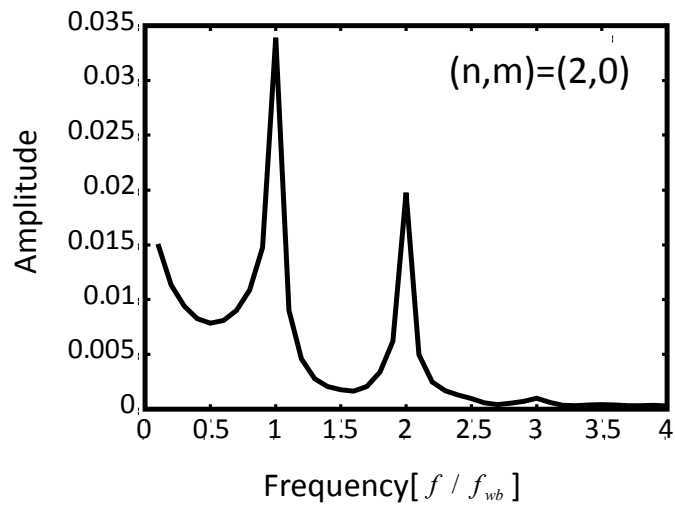

Fig. 9　Spectrum of the mode (2, 0) in its frequency space. The frequency $f_{wb}$ is the wobbling HIB frequency. The small nonuniformity of the HIBs energy deposition has the same frequency $f_{wb}$ of the wobblers and also the double frequency $2f_{wb}$.

When each HIB deposits its energy on a fuel target with the function of $e_i(\vec{r}, t)(1+\delta_i(\vec{r})e^{i\Omega t})$, the total HIBs energy deposition $E(\vec{r}, t)$ can be expressed as $E(\vec{r}, t) = \sum_{i=1}^{N_b} e_i(\vec{r}, t)(1+\delta_i(\vec{r})e^{i\Omega t}) = \sum_{i=1}^{N_b} e_i(\vec{r}, t) + \{\sum_{i=1}^{N_b} e_i(\vec{r}, t)\delta_i(\vec{r})\}e^{i\Omega t}$. Here $\delta_i(\vec{r})e^{i\Omega t}$ is the oscillating part of the deposited energy, $\Omega = 2\pi f_{wb}$, $\delta_i \ll 1$, $N_b$ is the total HIBs number, and in our case $N_b$=32. Therefore, the wobbling beams induce naturally the deposition energy oscillation with the wobbling oscillation frequency of $\Omega$.

The result in Fig. 9 demonstrates that the small nonuniformity of the HIBs energy deposition has the oscillation with the same frequency and the double frequency with the wobbling HIBs oscillation frequency of $f_{wb}$. In addition, the total nonuniformity magnitude is suppressed less than 3.87%. The results in this paper show a possibility of a rather uniform implosion in heavy ion fusion based on the wobbling HIBs. When a perturbation inducing the implosion nonuniformity emerges during the fuel target implosion based on the origin of the deposition energy nonuniformity, the perturbation may grow from the energy deposition nonuniformity with a certain phase. In the wobbling HIBs illumination, the nonuniformity phase is defined by the wobbling beams' motion. The overall perturbation growth is the superposition of all the perturbations with the different initial phases but with the same wavelength. Based on these considerations, it would be pointed out that the wobbling HIBs may contribute to a fuel target uniform implosion in heavy ion inertial fusion. [15, 16, 27]

In the analyses presented above, the HIB radius was fixed to be 3mm. In the wobbling HIBs illumination, the peak of the HIB energy deposition nonuniformity appears in the first few rotations. When the beam radius changes from 3.1mm to 3mm at $t = 1.3\tau$ during the spiral motion in the second rotations, we have an additional improvement for the HIBs illumination nonuniformity. The peak value of the nonuniformity becomes less than 3.57% (see Fig. 10), although the peak value of the nonuniformity in Fig. 6 was about 3.87%.

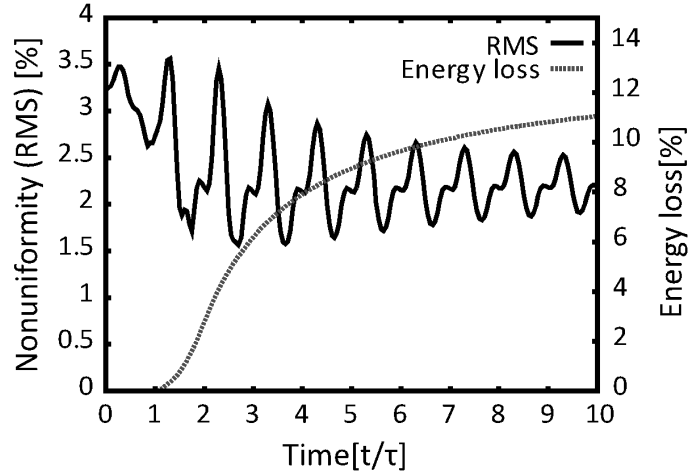

Fig. 10  Histories of the spiraling HIBs illumination nonuniformity (the solid line) and the illumination loss (the dotted line). The spiraling HIB radius changes at $t=1.3t$ from the initial beam radius of 3.1mm to 3.0mm.

## VI. ROBUST HIB ILLUMINATION

In this section a robustness of the spiraling HIBs' illumination scheme is examined against the target alignment error in a fusion reactor. We simulate the effect of a little displacement $dz$ as well as $dx$ and $dy$ on the HIB illumination nonuniformity as shown in Fig.

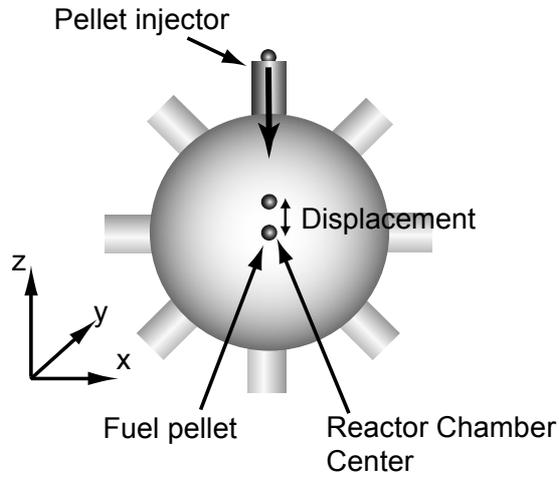

Fig. 11 Traget alignment error in a fusion reactor.

11. Figure 12 shows the relation between the pellet displacement and the maximal *rms* nonuniformity in the best case, in which each HIB radius changes from 3.1 mm to 3.0 mm at $t = 1.3\tau$. The maximal *rms* nonuniformity is less than 4.5 % for the pellet displacement less than about 80 μm. Figure 13 presents the spectrum of the mode (2, 0) amplitude in its frequency space.

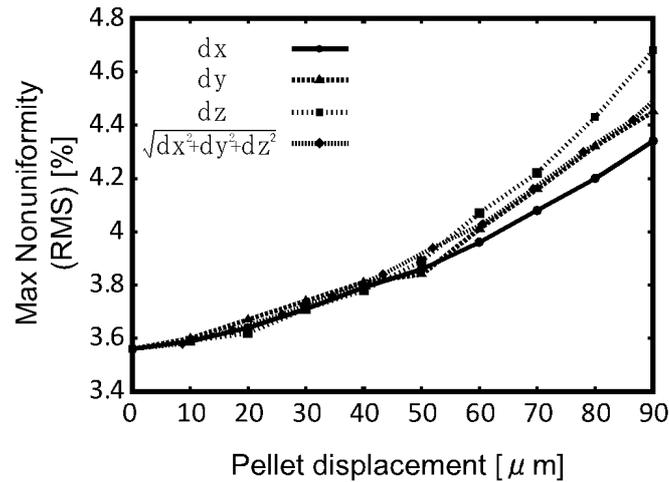

Fig. 12 The relation between the pellet displacement and the maximal rms nonuniformity in the best case, in which each HIB radius changes from 3.1 mm to 3.0 mm at $t = 1.3\tau$. The maximal rms nonuniformity is less than 4.5 % for the pellet displacement less than about 80 μm.

In the conventional beam illumination scheme, a fuel target displacement of 50-100 μm is tolerable.[20] The result shown in Fig. 12 means that the spiraling HIBs' illumination scheme provides the same order of the tolerable displacement with the conventional one. An enhancement of the allowable range[20] should be studied further in the future, if required.

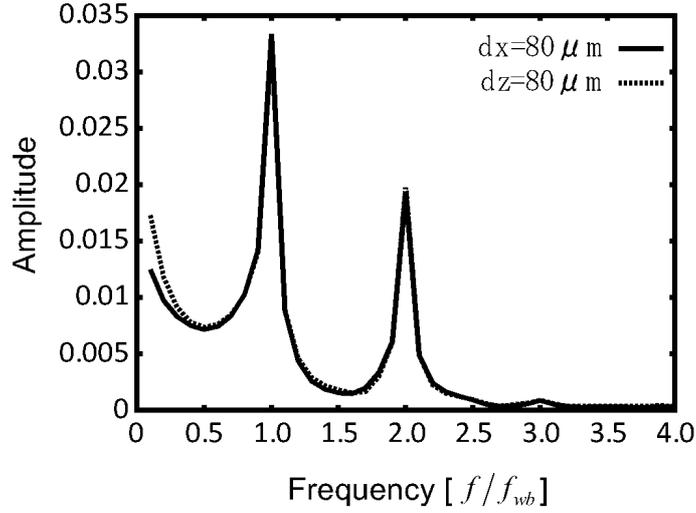

Fig. 13 The spectrum of the mode (2, 0) amplitude in its frequency space for *dx*=80mm and for *dz*=80mm.

## V. CONCLUSIONS

In this paper, we presented the HIBs illumination scheme of the spiraling HIBs, so that the HIBs illumination uniformity is drastically improved. Throughout the HIB input pulse, the beam illumination nonuniformity is kept low, that is, less than about 3.57%. The sufficiently small illumination nonuniformity is successfully realized by the wobbling HIBs onto a spherical target. The small-amplitude nonuniformity oscillation frequency is mostly the wobbling oscillation frequency and also the double of wobbling frequency. The wobbling HIBs may supply a viable uniform implosion mode in heavy ion fusion.

When the HIB input pulse is long so that HIBs radius zooming is needed,[28] we may also employ the inversely-spiraling illumination scheme to realize the zooming HIBs illumination to implode a converging compressed fuel with a low illumination nonuniformity.

**ACKNOWLEDGEMENTS**

This work is partly supported by MEXT, JSPS, ILE/Osaka Univ. and CORE (Center for Optical Research and Education, Utsunomiya university, Japan). The works is also partly supported by the Japan/U.S. Fusion Research Program.